\documentclass[prd,twocolumn,floatfix,nofootinbib,amsmath,amssymb,floatfix]{revtex4} 
\usepackage{graphicx,color,dcolumn,booktabs,bm,mathrsfs}
\usepackage{longtable,lscape}
\usepackage{txfonts}
\usepackage{overpic}
\usepackage{amssymb}
\usepackage{indentfirst}
\usepackage{feynmf}   
\usepackage{slashed}  
\usepackage{cases}
\usepackage{color}
\usepackage{multirow}
\usepackage{epstopdf}
\usepackage{CJK}

\def\tcsbar{T_{c\bar{s}0}(2900)}
\usepackage[colorlinks,
            citecolor=blue,
            anchorcolor=red,
            menucolor=red,
            linkcolor=red,
            filecolor=red,
            runcolor=red,
            urlcolor=blue,
            frenchlinks=red]{hyperref}
\usepackage{subfigure}

\begin{document}

\title{Searching for the open flavor tetraquark $T^{++}_{c\bar{s}0}(2900)$ in the process $B^+\to K^+ D^+ D^-$}

\author{Man-Yu Duan$^{1}$}
\author{En Wang$^{2}$}\email{wangen@zzu.edu.cn}
\author{Dian-Yong Chen$^{1,3}$\footnote{Corresponding author}}\email{chendy@seu.edu.cn}

\affiliation{$^1$School of Physics, Southeast University, Nanjing 210094, China\\
$^2$School of Physics and Microelectronics, Zhengzhou University, Zhengzhou, Henan 450001, China\\
$^3$Lanzhou Center for Theoretical Physics, Lanzhou University, Lanzhou 730000, P. R. China}

\begin{abstract}
Inspired by recent observations of $T_{c\bar{s}0}(2900)^0$ in the $D_s^+ \pi^-$ invariant mass distribution of $B^0 \to \bar{D}^0 D_s^+ \pi^-$ decay and $T_{c\bar{s}0}(2900)^{++}$ in the $D_s^+ \pi^+$ invariant mass distribution of $B^+ \to D^- D_s^+ \pi^+$ decay, we investigate the $\tcsbar^{++}$ contribution to the $B^+ \to K^+ D^+ D^-$ decay in a molecular scenario, where we consider $\tcsbar^{++}$ as a $D^{\ast +} K^{\ast+}$ molecular state. Our estimations indicate that the fit fraction of $\tcsbar^{++}$ in the $B^+ \to K^+ D^+ D^-$ is about $12.5\%$, and its signal is visible in the $D^+ K^+$ invariant mass distribution. With the involvement of $\tcsbar^{++}$, the fit fractions of $\chi_{c0}(3915)$ and $\chi_{c2}(3930)$ may be much different with the ones obtained by the present amplitude analysis [Phys. Rev. D \textbf{102}, 112003 (2020)], which may shed light on the long standing puzzle of $\chi_{c0}(3915)$ as the conventional charmonium.
\end{abstract}


\maketitle


\section{INTRODUCTION}
\label{sec:INTRODUCTION}


The $B$ meson decay process is the most productive and important platform of searching for the QCD exotic states. Two typical types of exotic candidates could be observed in this process. One is the charmonium-like state observed in the invariant mass distributions of a charmonium plus one or more light meson, such as the first observed charmonium-like state, $X(3872)$, which was first observed in the $\pi^+ \pi^- J/\psi$ invariant mass distribution of the process $B^\pm \to K^\pm \pi^+ \pi^- J/\psi$ by the Belle Collaboration in the year of 2003~\cite{Belle:2003nnu}, and then confirmed by the BaBar~\cite{Aubert:2004fc,Aubert:2004ns,Aubert:2005eg,Aubert:2005zh,Aubert:2005vi,Aubert:2006aj,Aubert:2007rva,Aubert:2008gu,Aubert:2008ae,delAmoSanchez:2010jr}, CDF~\cite{Acosta:2003zx,Abulencia:2005zc,Abulencia:2006ma,Aaltonen:2009vj}, D0~\cite{Abazov:2004kp}, CMS~\cite{CMS:2011yra,Vesentini:2012lea,Chatrchyan:2013cld,DallOsso:2013rtt,DallOsso:2014cmg,Sirunyan:2020qir}, and LHCb~\cite{Aaij:2011sn,LHCb:2011bia,LHCb:2011cra,Aaij:2013zoa,Aaij:2013rha,Aaij:2014ala,Aaij:2015eva,Aaij:2016kxn,Aaij:2017tzn,Aaij:2019zkm,Durham:2020zuw,Aaij:2020qga, Aaij:2020xjx,Aaij:2020tzn} in the $B$ decay process, as well as the BESIII~\cite{Ablikim:2013dyn,Ablikim:2019soz,Ablikim:2019zio,Ablikim:2020xpq} Collaboration in the electron-positron annihilation process. Besides the charmonium-like states, another type of exotic candidates observed in the $B$ decay processes is the open-charm states with strangeness observed in the invariant mass spectra of a charmed meson and a (anti-)kaon meson or $D_s \pi$, such as $D_{s0}^\ast (2317)$ and $D_{s1}(2460)$, which were first observed by BaBar \cite{BaBar:2003oey} and CLEO~\cite{CLEO:2003ggt} Collaborations, respectively.

In the year of 2020, the LHCb Collaboration performed the amplitude analysis of the process $B^+ \to D^- D^+ K^+$~\cite{LHCb:2020bls,LHCb:2020pxc}, and two new structures with spin-0 (named $X_0(2900)$) and spin-1 (named $X_1(2900)$), were reported in the $D^- K^+$ invariant mass distribution. The masses and widths of these two states are measured to be~\cite{LHCb:2020pxc,LHCb:2020bls}
\begin{eqnarray}
m_{X_0(2900)}&=&(2866\pm7\pm2)~\mathrm{MeV}\ ,\nonumber\\
\Gamma_{X_0(2900)}&=&(57\pm12\pm4)~\mathrm{MeV}\ ,\nonumber\\
m_{X_1(2900)}&=&(2904\pm5\pm1)~\mathrm{MeV}\ ,\nonumber\\
\Gamma_{X_1(2900)}&=&(110\pm11\pm4)~\mathrm{MeV}\ ,
\end{eqnarray}
respectively. 

It is interesting to notice that both $X_0(2900)$ and $X_1(2900)$ are fully open-flavor states and their minimal quark components are $\bar{c}\bar{s} u d$ , which indicates that $X_0(2900)$ and $X_1(2900)$ could be good candidates of tetraquark states~\cite{Agaev:2022eeh,Ozdem:2022ydv,Agaev:2021knl,Xue:2020vtq,Wang:2020xyc,Zhang:2020oze,He:2020jna,Lu:2020qmp}. In addition, the observed masses of $X_0(2900)$ and $X_1(2900)$ are close to the threshold of $D^\ast \bar{K}^\ast$, then the $D^\ast \bar{K}^\ast$ molecular interpretations have been proposed~\cite{Ke:2022ocs,Bayar:2022wbx,Chen:2021tad,Wang:2021lwy,Kong:2021ohg,Xiao:2020ltm,Agaev:2020nrc,Hu:2020mxp,He:2020btl,Liu:2020nil}.


Recently, the LHCb Collaboration reported two new tetraquark states $T_{c\bar{s}0}(2900)^0$ and $T_{c\bar{s}0}(2900)^{++}$ in the $D_s^+ \pi^-$ and $D_s^+ \pi^+$ mass distributions   of the $B^0 \to \bar{D}^0 D_s^+ \pi^-$ and $B^+ \to D^- D_s^+ \pi^+$, respectively~\cite{LHCb:2022xob,LHCb:2022bkt}. The masses and widths of the $T_{c\bar{s}0}(2900)^0$ and $T_{c\bar{s}0}(2900)^{++}$ are measured to be~\cite{LHCb:2022xob,LHCb:2022bkt}
\begin{eqnarray}
m_{T_{c\bar{s}0}(2900)^{0}}&=&(2892\pm14\pm15)~\mathrm{MeV}\ ,\nonumber\\
\Gamma_{T_{c\bar{s}0}(2900)^{0}}&=&(119\pm26\pm12)~\mathrm{MeV}\ ,\nonumber\\
m_{T_{c\bar{s}0}(2900)^{++}}&=&(2921\pm17\pm19)~\mathrm{MeV}\ ,\nonumber\\
\Gamma_{T_{c\bar{s}0}(2900)^{++}}&=&(137\pm32\pm14)~\mathrm{MeV}\ .
\end{eqnarray}
The resonance parameters of these two states are consistent with each other, which indicates that they are two of isospin triplet. When taking the isospin relationship into consideration, the mass and width of $T_{c\bar{s}0}(2900)$ are fitted to be~\cite{LHCb:2022xob,LHCb:2022bkt},
\begin{eqnarray}
m_{T_{c\bar{s}0}(2900)}&=&(2908\pm11\pm20)~\mathrm{MeV}\ ,\nonumber\\
\Gamma_{T_{c\bar{s}0}(2900)}&=&(136\pm23\pm11)~\mathrm{MeV}\ .
\end{eqnarray}
In addition, the amplitude analysis indicates the quantum numbers of $T_{c\bar{s}0}$ are $J^P=0^+$. 

From the observed processes, one can find the minimal quark components of $T_{c\bar{s}0}(2900)^0$ and $T_{c\bar{s}0}(2900)^{++}$ are $c\bar{s} \bar{u} d$ and $c\bar{s} \bar{d} u$, respectively, which indicates that both $T_{c\bar{s}0}(2900)^0$ and $T_{c\bar{s}0}(2900)^{++}$ are also fully open flavor tetraquark states, and in addition, $T_{c\bar{s}0}(2900)^{++}$ is the first observed doubly charged tetraquark state. These particular properties have stimulated theorists' great interests. In the framework of the QCD sum rules, the authors in Ref.~\cite{Yang:2023evp, Lian:2023cgs, Jiang:2023rcn, Liu:2022hbk, Dmitrasinovic:2023eei} assigned $T_{c\bar{s}0}(2900)$ as the scalar $c\bar{s}q\bar{q}$ tetraquark state.  In addition, the observed mass of $T_{c\bar{s}0}(2900)$ is close to the threshold of $D^\ast K^\ast$. Together with $D_{s0}^*(2317)$ close to the $DK$ threshold and $D_{s1}(2460)$ close to the $D^\ast K$ threshold, the observation of $T_{c\bar{s}0}(2900)$ enrich the exotic candidate near the threshold of a charmed meson and a strange meson. Similar to the case of $D_{s0}^*(2317)$ and $D_{s1}(2460)$,  $T_{c\bar{s}0}(2900)$ has also been proposed to be $D^\ast K^\ast$ molecular state with isospin $I=1$. By means of the QCD two-point sum rule method, the mass and decay width could be reproduced in the $D^\ast K^\ast$ molecular scenario~\cite{Agaev:2022eyk}. In the one-boson-exchange model, the authors in Ref.~\cite{Chen:2022svh} found that the masses of $D_{s0}^{\ast}(2317)$, $D_{s1}(2460)$ and $T_{c\bar{s}0}(2900)$ could be reproduced. In an effective Lagrangian approach, the decay properties of $T_{c\bar{s}0}(2900)$ were also investigated in Ref.~\cite{Yue:2022mnf}. Besides the resonance interpretations, the $T_{c\bar{s}0}(2900)$ was interpreted as the threshold effect from the interaction of the $D^{\ast}K^{\ast}$ and $D_s^{\ast}\rho$ channels~\cite{Molina:2022jcd} or the triangle singularity~\cite{Ge:2022dsp}.

On the experimental side, searching for more decay modes of $T_{c\bar{s}0}(2900)$ can help us to reveal its internal structure. In the $B^+ \to D^- D^+ K^+$ process where the tetraquark states $X_0(2900)$ and $X_1(2900)$ were observed, the LHCb Collaboration also present  the $D^+ K^+$ invariant mass distribution~\cite{LHCb:2020bls,LHCb:2020pxc}. From the measured data, one find that the $D^+ K^+$ invariant mass distribution can not be well described in the vicinity of 2.9 GeV\footnote{More detail can be found in Fig.10-(c) of Ref.~\cite{LHCb:2020pxc}}, which indicates that there could be some contributions from additional resonances. To further analyse the resonance contributions to $B^+ \to D^- D^+ K^+$ process, we find, 
\begin{itemize}
\item Besides the resonance parameters of $T_{c\bar{s}0}(2900)$,  the LHCb Collaboration also reported the fit fraction of $T_{c\bar{s}0}(2900)^{++}$ component in the $B^+ \to D^- D_s^+ \pi^+$, which is $(1.96\pm 0.87 \pm 0.88)\%$~\cite{LHCb:2022xob,LHCb:2022bkt}. In other words, the cascaded decay process, $B^+ \to D^- T_{c\bar{s}0}(2900)^{++} \to D^- D_s^+ \pi^+$ are sizable. 
\item In the $D^\ast K^\ast$ molecular scenario, the decay properties of the $T_{c\bar{s}0}(2900)^0$ were investigated in Ref.~\cite{Yue:2022mnf}. Our estimations indicate that the $T_{c\bar{s}0}(2900)^0$ dominantly decays into $D^0 K^0$, and accordingly $T_{c\bar{s}0}(2900)^{++}$ should dominantly decay into $D^+ K^+$ on account of the isospin symmetry.  
\end{itemize} 
Based on the above experimental measurements and theoretical estimations, one can anticipate that the tetraquark state $T_{c\bar{s}0}(2900)^{++}$ should have non-negligible contribution to the process  $B^+ \to  D^- (K^+ D^+)$. 

In addition, the involvement of $\tcsbar^{++}$ in the process $B^+\to K^+ D^+ D^-$ may also shed light on another long standing puzzle for $\chi_{c0}(3930)$ as conventional charmonium~\cite{Zhou:2015uva, Chen:2012wy, Olsen:2014maa, Guo:2012tv}. The measurements from the BaBar Collaboration indicated that the branching fraction of $B^+\to K^+ \chi_{c0}(3930) \to K^+ J/\psi \omega$ is $(3.0^{+0.7+0.5}_{-0.6-0.3})\times 10^{-5}$~\cite{BaBar:2010wfc}, while the branching fraction of $B^+ \to K^+ \chi_{c0}(3930) \to K^+ D^+ D^-$ is reported to be $(8.1\pm 3.3)\times 10^{-6}$~\cite{LHCb:2020pxc}. Thus, one can conclude that the branching fraction for $\chi_{c0}(3930)\to J/\psi \omega$ is several times larger than the one of $\chi_{c0}(3930) \to D^+ D^-$, which is inconsistent with the expectations of the conventional charmonium assignment of $\chi_{c0}(3930)$.

If carefully checking the $D^+ K^+$ invariant mass distribution of $B^+ \to K^+ D^+ D^-$ in Ref.~\cite{LHCb:2020pxc}, one can find that the charmonium $\chi_{c2}(3930)$ has significant contribution to the structure near 2.9 GeV in the $D^+K^+$ invariant mass distribution. While both the $\chi_{c0}(3930)$ and $\chi_{c2}(3930)$ are responsible for the peak in the vicinity of 3.93 GeV in the $D^+ D^-$ mass spectrum of $B^ +\to K^+ D^+ D^-$, then, the  involvement of $\tcsbar^{++}$ in the $B^+ \to K^+ D^+ D^-$ may lead to a rather different fit fractions of $\chi_{c0}(3930)$ and $\chi_{c2}(3930)$ with the present one. Thus, in the present work, we investigate the possible contribution of $T_{c\bar{s}0}(2900)^{++}$ in the process $B^+ \to D^- D^+ K^+$ in the framework of the molecular scenario, where $\tcsbar^{++}$ is considered as a $D^{\ast+} K^{\ast +} $ molecular state.

This paper is organized as follows. After the introduction, we will show the formalism used in Sec.~\ref{sec:FORMALIISM}. Our calculated results and related discussions will be presented in Sec.~\ref{sec:RESULTS}, and Sec.~\ref{sec:Summary} will devote to a short summary.


\begin{figure}[htb]
  \centering
  \subfigure{\includegraphics[scale=0.55]{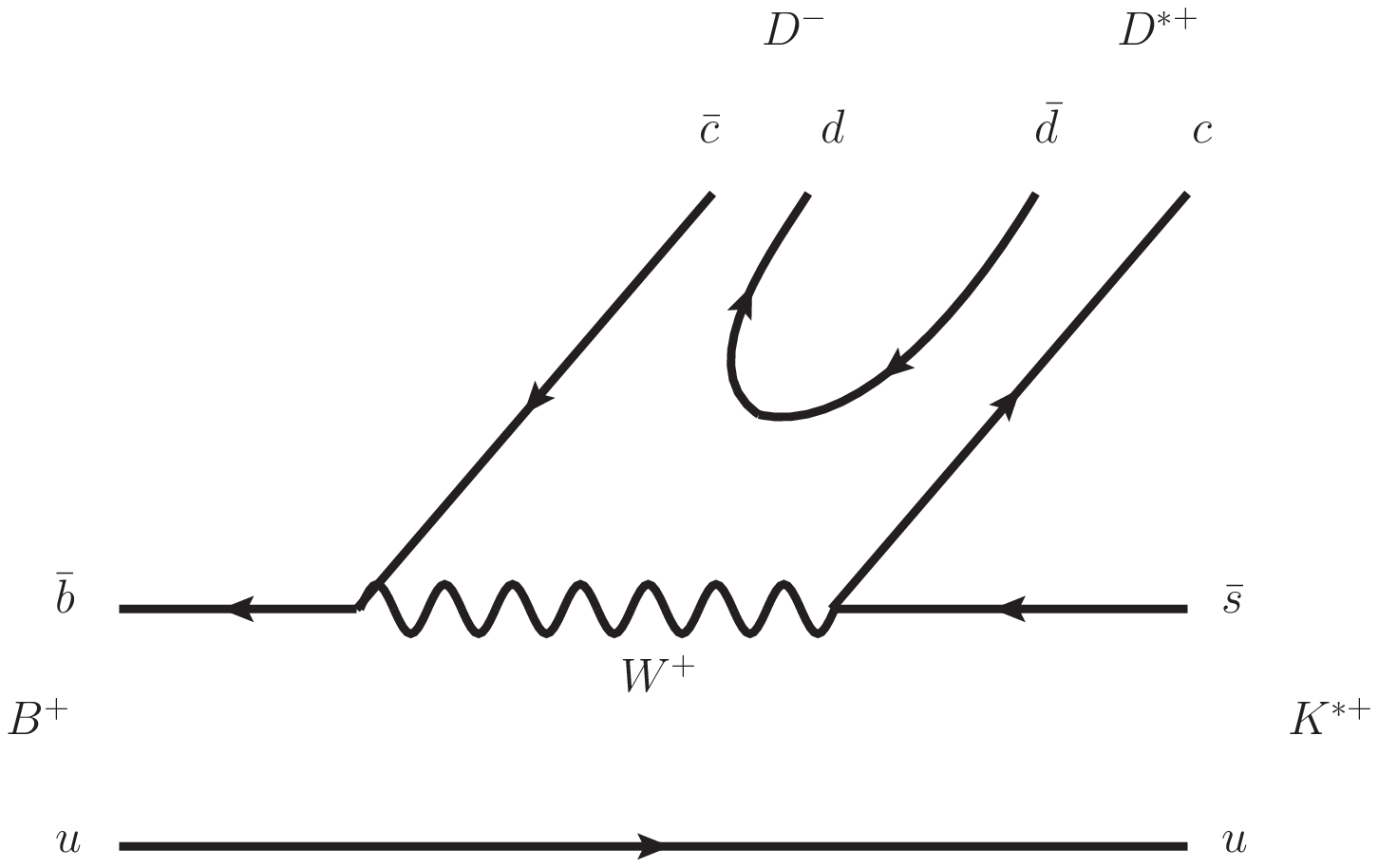}\label{Bp.eps}}
  \caption{Diagrammatic decay at the quark level for the $B^+ \to D^- D^{\ast +} K^{\ast+}$ reaction.}
  \label{Fig:Mec0}
  \end{figure}

\begin{figure*}[thb]
  \centering
  \subfigure{\includegraphics[scale=0.55]{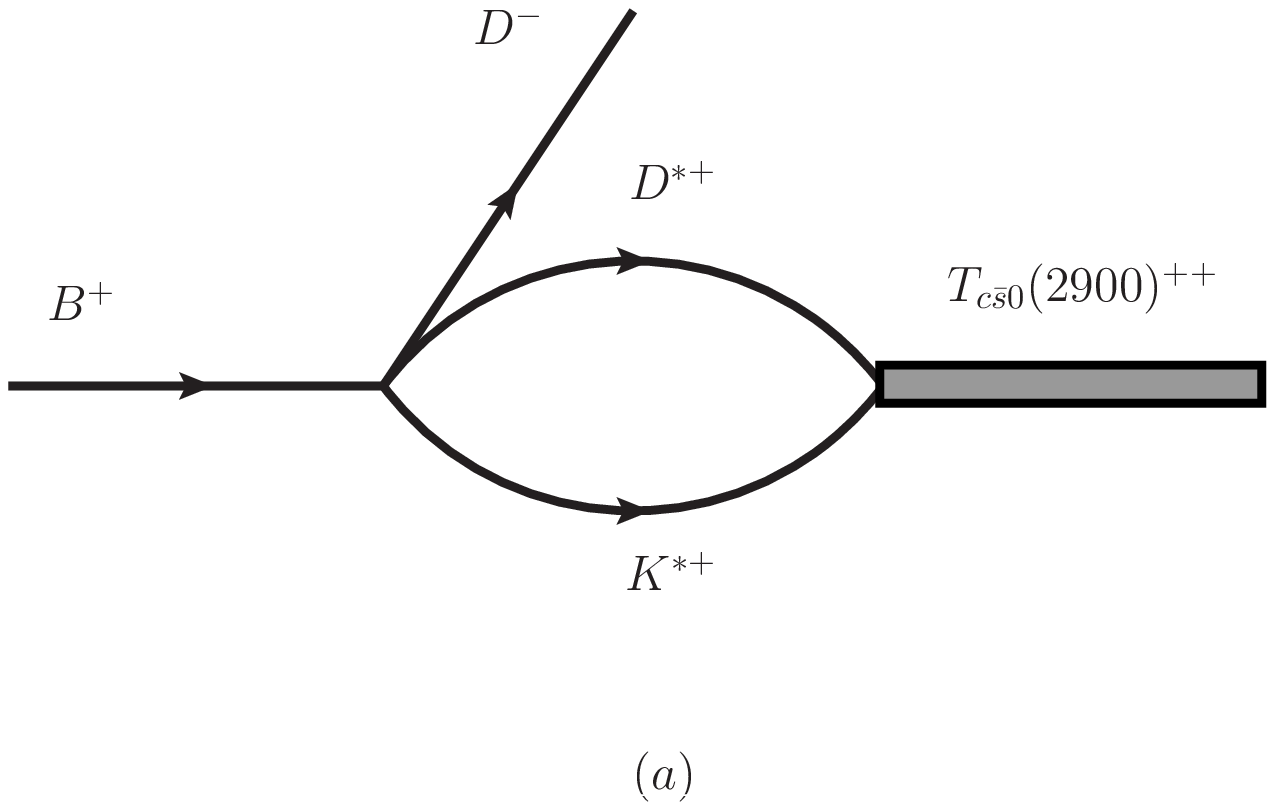}\label{pp1.eps}}
  \subfigure{\includegraphics[scale=0.55]{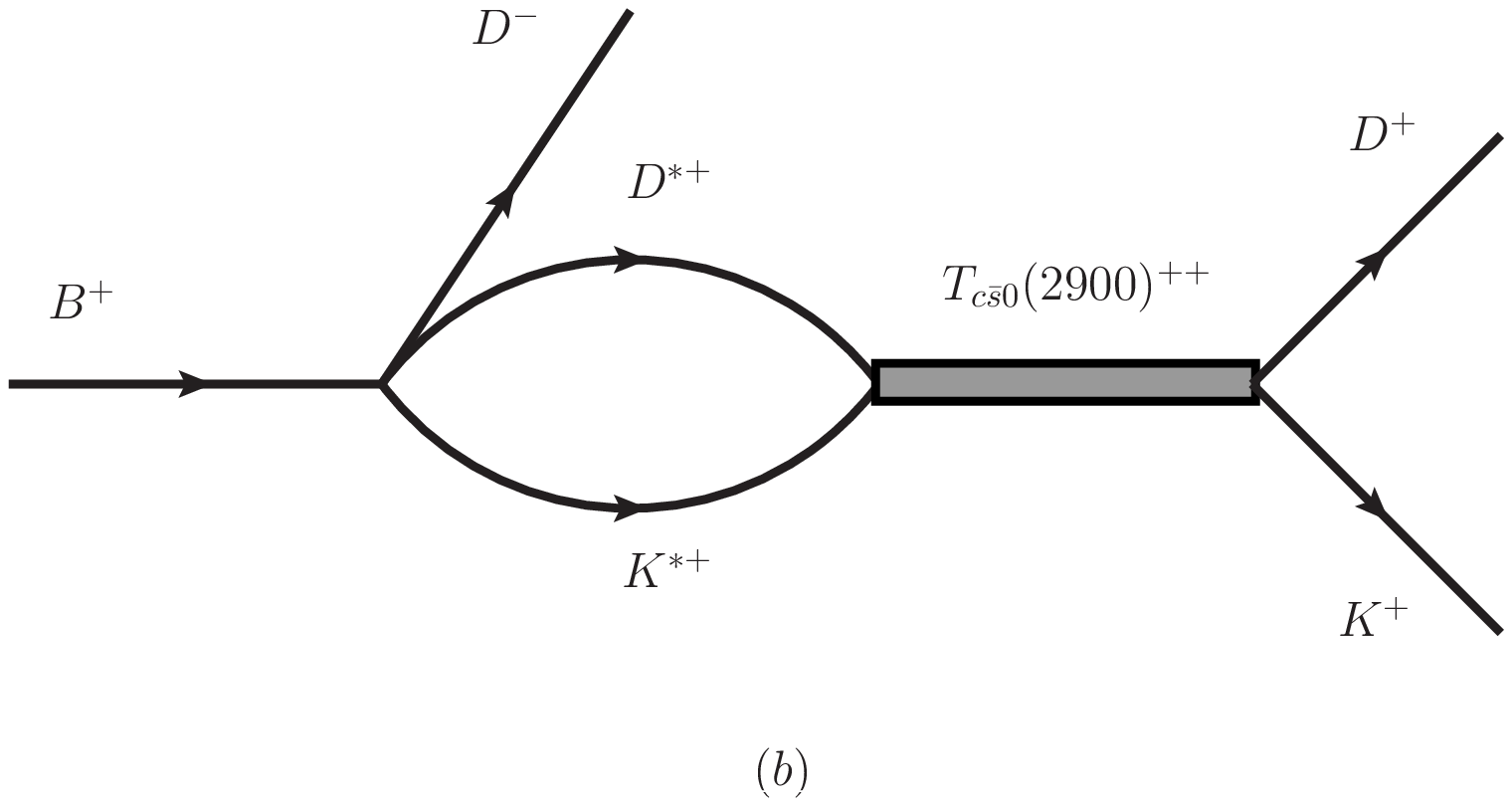}\label{pp.eps}}
  \caption{A sketch diagram of the rescatting of $D^{\ast+} K^{\ast+}$ to give the resonance $T_{c\bar{s}0}(2900)^{++}$ (diagram (a)), and further decay of $T_{c\bar{s}0}(2900)^{++}$ to $D^+ K^+$ (diagram (b)).}
  \label{Fig:Mec1}
  \end{figure*}

 \section{FORMALISM}
\label{sec:FORMALIISM}
In the molecular scenario, the $\tcsbar^{++}$ is considered as a molecular composed of $D^{\ast+} K^{\ast+}$,  which is,
\begin{eqnarray}
	\left|\tcsbar^{++}\rangle \right.= \left| D^{\ast +} K^{\ast +}\rangle \right. .
\end{eqnarray}
Thus, the primary reaction that could produce $\tcsbar^{++}$ is $B^+ \to D^- D^{\ast +} K^{\ast+}$. As shown in Fig.~\ref{Fig:Mec0}, this reaction proceeds via the $W^+$ internal emission, where the $\bar{b}$ quark transits into $\bar{c}$ quark by emitting a $W^+$ boson, while the $W^+$ boson couples to the $c\bar{s}$ quarks pair. The $\bar{s}$ quark and the $u$ quark from the initial $B^+$ meson form a $K^{\ast+}$ meson, while the rest $c\bar{c}$ and $d\bar{d}$ created from vacuum hadronize  into $D^{-}$ and $D^{\ast +}$ mesons. In the hadron level, one can construct the $S$ wave component of the transition amplitude by matching the angular momentum of $B^+$ meson~\cite{Dai:2022htx,Dai:2022qwh}, which is,
\begin{eqnarray}
-i t_1 = -i C_1 {\bm \epsilon (D^{*+} ) } \cdot {\bm \epsilon (K^{*+} ) },
\label{eq:tildet}
\end{eqnarray}
where the $\epsilon(D^{\ast+} )$ and $\epsilon(K^{\ast +})$ are the polarization vectors of the $D^{\ast+}$ and $K^{\ast +}$, respectively. $C_1$ is an unknown coupling constant, which will be  discussed later.  Then the $D^{\ast+}$ and $K^{\ast +}$ couple to the molecular $\tcsbar^{++}$ with $I(J^P)=1(0^+)$ as presented in Fig.~\ref{Fig:Mec1}-(a). As indicated in Ref.~\cite{Molina:2008jw}, the spin of the $D^{\ast+} K^{\ast+}$ system could be projected into different angular momentum, for example, the vertex for $R_{J}\to D^{\ast+} K^{\ast+}$ with $J=0,1,2$ could be constructed as,
\begin{eqnarray}
\mathcal{V}^{(0)}&=& \frac{1}{3} \epsilon_l(D^{*+}) \epsilon_l(K^{*+}) \delta_{ij} , \nonumber \\
\mathcal{V}^{(1)}&=& \frac{1}{2} \left [ \epsilon_i(D^{*+}) \epsilon_j(K^{*+}) -\epsilon_j(D^{*+}) \epsilon_i(K^{*+}) \right ] ,  \nonumber  \\
\mathcal{V}^{(2)}&=& \frac{1}{2} \left [ \epsilon_i(D^{*+}) \epsilon_j(K^{*+}) +\epsilon_j(D^{*+}) \epsilon_i(K^{*+}) \right ]   \nonumber \\
& &- \frac{1}{3} \epsilon_l(D^{*+}) \epsilon_l(K^{*+}) \delta_{ij}.
\label{Eq:Project}
\end{eqnarray}
The experimental analysis indicated that the angular momentum of $\tcsbar$ is $0$. Thus, one can obtain the transition amplitude of $B^+ \to D^-  \tcsbar^{++}$ corresponding to Fig.~\ref{Fig:Mec1}-(a), which is,
\begin{eqnarray}
-i t_{2a} &=& -i  C_1 \epsilon_{\alpha} (D^{*+} )  \epsilon_{\beta} (K^{*+} ) \delta^{\alpha \beta} G_{D^* K^*} (M_{\rm inv}(D^+K^+))  \nonumber \\
& & \times  \frac{1}{3} \epsilon_l^* (D^{*+} )  \epsilon_l^* (K^{*+} ) \delta_{ij} g_{T^{++}_{c\bar{s}0}D^* K^*}  \nonumber \\
&=& -i C_1 \delta_{ij} G_{D^* K^*} (M_{\rm inv}(D^+K^+)) g_{T^{++}_{c\bar{s}0}D^* K^*},
\end{eqnarray}
where $\sum\limits_{pol} \epsilon_i (R) \epsilon_j^* (R) = \delta_{ij}$, $R = D^{*+}$ or $K^{*+}$, and the sum over the same indices of the Kronecker delta function is equal to 3, i.e., $\sum\limits_{ij} |\delta_{ij}|^2 = 3$. $G_{D^* K^*} (M_{T_{c\bar{s}0}})$ is the loop function of the two-meson $D^\ast$ and $K^\ast$, which will be discussed later.

Similarly, one can obtain the transition amplitude of $B^+ \to D^- T_{c\bar{s}0}(2900)^{++} \to D^- D^+ K^+$ corresponding to Fig.~\ref{Fig:Mec1}-(b), which is,
\begin{eqnarray}
-it_{2b} &=& -i C_1 \delta_{ij} G_{D^* K^*} \left(M_{\rm inv}(D^+K^+)\right) \nonumber\\ &&\times \frac{g_{T^{++}_{c\bar{s}0}D^* K^*} g_{T^{++}_{c\bar{s}0}D K}}{M^2_{\rm inv}(D^+K^+) -m^2_{T^{++}_{c\bar{s}0}}+i m_{T^{++}_{c\bar{s}0}}\Gamma_{T^{++}_{c\bar{s}0}}} ,
\end{eqnarray}  
and then the square of the transition amplitude is,
\begin{eqnarray}
\sum |t_{2b}|^2 &=& 3 C_1^2 \  \Big|G_{D^* K^*} \left(M_{\rm inv}(D^+K^+)\right)\Big|^2\     \nonumber\\
& &\times \frac{\Big|g_{T^{++}_{c\bar{s}0},D^* K^*}\Big|^2 \Big|g_{T^{++}_{c\bar{s}0},D K}\Big|^2}{\left[M^2_{\rm inv}(D^+K^+)-m^2_{T^{++}_{c\bar{s}0}}\right]^2+ m_{T^{++}_{c\bar{s}0}}^2 \Gamma^2_{T^{++}_{c\bar{s}0}}},
\label{eq:t1}
\end{eqnarray}
with $M^2_{\rm inv}(D^+K^+)=(P_{D^+}+P_{K^+})^2$, and two-meson loop function is given by,
\begin{equation}
G=i\int\frac{d^4q}{(2\pi)^4}\frac{1}{q^2-m_1^2+i\epsilon}\frac{1}{(q-P)^2-m_2^2+i\epsilon}\ ,
\label{eq:loopex}
\end{equation}
with $m_1$ and $m_2$ the masses of the two mesons involved in the loop. $q$ is the four-momentum of the meson in the centre of mass frame, and $P$ is the total four-momentum of the meson-meson system. In the present work, we use the dimensional regularization method as indicated in Refs.~\cite{Duan:2022upr,Duan:2021pll,Duan:2020vye},  and in this scheme, the two-meson loop function $G$ can be expressed as,
\begin{eqnarray}
G&=&\frac{1}{16\pi^2}\left[\alpha+\log\frac{m_1^2}{\mu^2}+\frac{m_2^2-m_1^2+s}{2s}\log\frac{m_2^2}{m_1^2}\right. \nonumber\\
&&+\frac{|\vec{q}\,|}{\sqrt{s}}\left(\log\frac{s-m_2^2+m_1^2+2|\vec{q}\,|\sqrt{s}}{-s+m_2^2-m_1^2+2|\vec{q}\,|\sqrt{s}}\right. \nonumber \\
&&+\left. \left. \log\frac{s+m_2^2-m_1^2+2|\vec{q}\,|\sqrt{s}}{-s-m_2^2+m_1^2+2|\vec{q}\,|\sqrt{s}}\right)\right] ,
\label{eq:loopexdm}
\end{eqnarray}
where $s=P^2=M^2_{\rm inv}(D^+K^+)$, and $\vec{q}\,$ is the three-momentum of the meson in the centre of mass frame, which reads,
\begin{equation}
|\vec{q}\,|=\frac{\sqrt{\left[s-(m_1+m_2)^2\right]\left[s-(m_1-m_2)^2\right]}}{2\sqrt{s}} ,
\end{equation}
here we take $\mu=1500$~MeV and $\alpha=-1.474$, which are the same as those in the study of the $D^*\bar{K}^*$ interaction~\cite{Dai:2022qwh, Dai:2022htx}. 

Besides the two-meson loop function, two coupling constants $g_{T^{++}_{c\bar{s}0},D^* K^*}$ and $g_{T^{++}_{c\bar{s}0},D K}$ are unknown. As for $g_{T^{++}_{c\bar{s}0},D^* K^*}$, it refers to the coupling between $\tcsbar^{++}$ and its components $D^{\ast+} K^{\ast+}$, which could be related to the binding energy by~\cite{Weinberg:1965zz, Baru:2003qq, Wu:2023fyh},
\begin{equation}
g_{T^{++}_{c\bar{s}0},D^* K^*}^2=16 \pi (m_{D^*}+m_{K^*})^2 \tilde{\lambda}^2 \sqrt{\frac{2 \Delta E}{\mu}} ,
\end{equation}
where $\tilde{\lambda} =1$ gives the probability to find the molecular component in the physical states, $\Delta E = m_{D^*}+m_{K^*} - m_{T^{++}_{c\bar{s}0}}$ denotes the binding energy, and $\mu = m_{D^*} m_{K^*} / (m_{D^*}+m_{K^*})$ is the reduced mass. 

As for $g_{T^{++}_{c\bar{s}0},D K}$, we tried to obtain its value by the corresponding partial width of $\tcsbar^{++} \to D^+ K^+$, with an effective Lagrangian approach, the partial width of $\tcsbar^{++} \to D^+ K^+$ could be obtained as,
\begin{eqnarray}
\Gamma_{T^{++}_{c\bar{s}0}}&=&\frac{1}{8\pi} \frac{1}{m^2_{T^{++}_{c\bar{s}0}}} |g_{T^{++}_{c\bar{s}0},D K}|^2 |\vec{q}_{K^+}| , 
\label{eq:coupling1}
\end{eqnarray}
with
\begin{eqnarray}
|\vec{q}_{K^+}|&=&\frac{\lambda^{1/2}(m^2_{T^{++}_{c\bar{s}0}},m^2_{D^+},m^2_{K^+})}{2 m_{T^{++}_{c\bar{s}0}}},
\end{eqnarray}
to be the momentum of $K^+$ in the $\tcsbar^{++}$ rest frame, and $\lambda(x,y,z)=x^2+y^2+z^2-2xy-2yz-2xz$ is the K$\ddot{\textrm{a}}$llen function. In Ref.~\cite{Yue:2022mnf}, our estimations indicated that the $\tcsbar^{++}$ dominantly decay into  $DK$, and the partial width of $DK$ channel was estimated to be $(52.6\sim 101.7)$ MeV in the considered parameter range. In the present work, we take the partial width of $\tcsbar^{++} \to D^+ K^+$ to be 80~MeV to estimate the coupling constant $g_{T_{c\bar{s}0}^{++} DK}$.

With the above preparation, one can obtain the $D^+ K^+$ invariant mass distribution, which is,
\begin{eqnarray}
\frac{d\Gamma}{dM_{\rm inv} (D^+ K^+)}=\frac{1}{(2\pi)^3} \frac{1}{4m^2_{B^+}} p_{D^-} \tilde{p}_{K^+} \sum |t_{2b}|^2 ,
\label{eq:inv1}
\end{eqnarray}
with
\begin{eqnarray}
 p_{{D}^-}&=&\frac{\lambda^{1/2}\left( m^2_{B^+},m^2_{D^-},M^2_{\rm inv}(D^+ K^+)\right)}{2 m_{B^+}} , \nonumber\\
 \tilde{p}_{K^+}&=&\frac{\lambda^{1/2}\left( M^2_{\rm inv}(D^+ K^+),m^2_{D^+},m^2_{K^+}\right)}{2 M_{\rm inv}(D^+ K^+)} .
\end{eqnarray}

In addition, we would like to compare the above mass distribution with the one of the background for the reaction $B^+ \to D^- D^+ K^+$. By analogy to Eq.~\eqref{eq:tildet}, we can obtain the transition matrix for $B^+ \to D^- D^+ K^+$, which is 
\begin{eqnarray}
-i t_3 = -i C_3.
\end{eqnarray}
where $C_3$ is the coupling constant, which will be discussed in the following section. With the above transition matrix, we can give the background distribution for the $B^+ \to D^- D^+ K^+$ reaction, which is
\begin{eqnarray}
\frac{d\Gamma_{\rm bac}}{dM_{\rm inv} (D^+ K^+)}=C_3^2 \frac{1}{(2\pi)^3} \frac{1}{4m^2_{B^+}} p_{D^-} \tilde{p}_{K^+} .
\label{eq:bac1}
\end{eqnarray}


\section{Numerical RESULTS AND DISCUSSIONS}
\label{sec:RESULTS}

To calculate the $D^{\ast+} K^{\ast+}$ invariant mass distribution of $B^+\to K^+ D^+ D^-$ as presented in Eq. (\ref{eq:inv1}), the coupling constant $C_1$ is needed. However, the experimental measurement of $B^+ \to K^{\ast +} D^{\ast+} D^{-}$ is not available to date. Similar to  $B^{+}\to K^+ D^+ D^-$, the process $B^+ \to K^{\ast +} D^{\ast+} D^{-}$ should also occur via $W^+$ internal  emission process. One can obtain the diagrammatic decay at the quark level for the $B^+ \to K^+ D^+ D^-$ by replacing $K^{\ast+}$ and $D^{\ast+}$ in Fig.~\ref{Fig:Mec0} with $K^+$ and $D^+$, which indicates some similarities between the processes $B^+ \to K^+ D^+ D^-$ and $B^+ \to K^{\ast +} D^{\ast+} D^{-}$. However, there are also some differences between these two processes. As indicated in the amplitude analysis of $B^+ \to K^+ D^+ D^-$ in Ref.~\cite{LHCb:2020pxc}, the typical resonance contributions to this process are $B^+ \to K^+ (c\bar{c})\to K^+ D^+ D^-$, where the charmonia include $\psi(3770)$, $\chi_{c0}(3930)$, $\chi_{c2}(3930)$, $\psi(4040)$, $\psi(4160)$, and $\psi(4415)$. These charmonia contributions should be suppressed due to phase space. In addition to the charmonia contributions, the LHCb Collaboration also observed the signals of $X_{0}(2900)$ and $X_1(2900)$ in the $D^- K^+$ invariant mass spectra, these contributions also vanish in the $B^+ \to K^{\ast +} D^{\ast+} D^{-}$ process.

Besides the resonance contributions, the amplitude analysis  also indicates sizable nonresonant contribution, which should be the same for both $B^+ \to K^+ D^+ D^-$ and $B^+ \to K^{\ast +} D^{\ast+} D^{-}$, thus, in the present work, we first estimate the background distribution of $B^+\to K^+ D^+ D^-$ with the branching fraction of the nonresonant contribution from LHCb analyze, which is $(5.3 \pm 1.8)\times 10^{-5}$~\cite{Workman:2022ynf}. From Eq. (\ref{eq:bac1}), the coupling constant $C_3$ could be determined. Considering the similarity between $B^+ \to K^+ D^+ D^-$ and $B^+ \to K^{\ast +} D^{\ast+} D^{-}$, we take $C_1=C_3$ to roughly estimate the $D^+ K^+$ invariant mass distribution resulted from $\tcsbar^{++}$.

  \begin{figure}[t]
  \centering
 \includegraphics[scale=1]{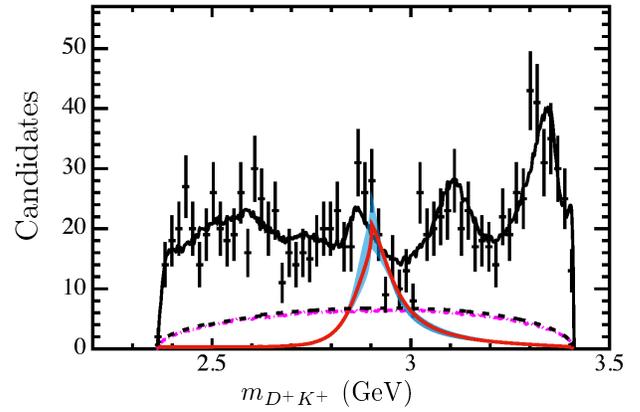}
  \caption{The $D^+ K^+$ invariant mass distribution for the $B^+ \to D^- D^+ K^+$ reaction.}
  \label{Fig:Tcs}
  \end{figure}
  
With the above formalism, we have calculated the $D^+ K^+$ invariant mass distribution by assuming the values of $C_1$ and $C_3$ are the same, as presented in Fig.~\ref{Fig:Tcs}. To further compare with the experimental measurements, we normalized the background contribution estimated by Eq. (\ref{eq:bac1}) to the LHCb experimental nonresonant contribution in Fig.~\ref{Fig:Tcs}, where the magenta-dash-dotted and blue-dotted curves are the nonresonant contribution determined by the LHCb amplitude analysis and our estimated  background, respectively. The red-solid curve is the resonant contribution form $\tcsbar^{++}$, which is obtained with the resonance parameters of $m_{T_{c\bar{s}0}^{++}}=2885 \ \mathrm{MeV}$ and $\Gamma_{T_{c\bar{s}0}^{++}}=136 \ \mathrm{MeV}$. While the blue band corresponds to the uncertainties of the $\tcsbar^{++}$ width. From Fig.~\ref{Fig:Tcs}, one can find that the $D^+K^+$ invariant mass distribution around 2.9~GeV can not be well described by LHCb fit~\cite{LHCb:2020pxc}, which indicates that there should be an additional resonance. Our results show that the $\tcsbar^{++}$ plays an important role in this region, thus we suggest that contribution from the $\tcsbar^{++}$ should be considered in the future amplitudes analysis.   

Furthermore, we can integrate the invariant mass $M_{\rm inv} (D^+ K^+)$ over the whole invariant mass range for the signal and background, and their ratio is given by, 
\begin{eqnarray}
\frac{\int\frac{d\Gamma}{dM_{\rm inv} (D^+ K^+)}}{\int\frac{d\Gamma_{bac}}{dM_{\rm inv} (D^+ K^+)}} \simeq 0.52.
\label{eq:vs1}
\end{eqnarray}
With the nonresonant fit fraction obtained by the amplitude analyze, we can roughly estimate the fit fraction of $\tcsbar^{++}$ to be about $12.5\%$, which is greater than the ones of $\chi_{c0}(3930)$ and $\chi_{c2}(3930)$. Thus, the involvement of $\tcsbar^{++}$ will certainly influence the fit fractions of $\chi_{c0}(3930)$ and $\chi_{c2}(3930)$.


\section{Summary}
\label{sec:Summary}

Recently, the LHCb Collaboration reported their amplitude analysis of the decays $B^0 \to \bar{D}^0 D_s^+ \pi^-$ and $B^+ \to D^- D_s^+ \pi^+$, where two tetraquark states $\tcsbar^{0}$ and $\tcsbar^{++}$ were reported in the $D_s \pi$ invariant mass distributions. The resonance parameters of these two resonances indicate that they are two of the isospin triplet. Similar to $\tcsbar$, the LHCb Collaboration reported another two tetraquark candidates $X_{0,1}(2900)$ in the $D^- K^+$ invariant mass distribution in the $B^+ \to D^- D^+ K^+$ reaction in the year of 2020 ~\cite{LHCb:2020bls,LHCb:2020pxc}. In the $D^+ K^+$ invariant mass distribution of the $B^+ \to D^- D^+ K^+$ reaction, we find that the experimental data of the $D^+K^+$ invarinat mass distribution around 2.9~GeV can not be well described, which indicates that there should be an additional resonance. Inspired by the recent observation of the $T_{c\bar{s}0}(2900)$~\cite{LHCb:2022xob,LHCb:2022bkt} and the decay properties of $T_{c\bar{s}0}(2900)$, we find that $\tcsbar^{++}$ is likely to contribute to the $D^+ K^+$ invariant mass distribution. Thus, in the  present work we study the role of $T_{c\bar{s}0}(2900)^{++}$ in the $D^+ K^+$ invariant mass distribution of the process $B^+ \to D^- D^+ K^+$.

In the present work, we estimate $T_{c\bar{s}0}(2900)^{++}$ contribution to the process $B^+ \to D^- D^+ K^+$ in a molecular scenario, where we have considered $\tcsbar^{++}$ as a $D^{\ast +} K^{\ast+}$ molecular state. However, due to the lack of the experimental information of $B^+ \to D^- D^{\ast +} K^{\ast+}$, we have made an assumption that the coupling constant for $B^+ \to D^- D^{\ast +} K^{\ast+}$ is the same as the one for nonresonant contribution in $B^+ \to D^- D^{ +} K^{+}$. Based on this assumption, our estimation indicates that the contribution from $\tcsbar^{++}$ is significant in the process $B^+ \to D^- D^+ K^+$, and the $\tcsbar^{++}$ signal in the $D^+ K^+$ invariant mass distribution is visible. In addition, the fit fraction of $B^+ \to D^- \tcsbar^{++}\to K^+ D^+ D^-$ is roughly estimated to be $12.5\%$, which could be tested by further experimental analysis by the LHCb Collaboration.

Before the end of this work, it is worth to mention that the branching fractions of $B^0 \to D^- D^0 K^+$ and  $B^0 \to D^- D^+ K^0$ decays are $( 1.07 \pm 0.07 \pm 0.09 ) \times 10^{-3}$ and $( 0.75 \pm 0.12 \pm 0.12 ) \times 10^{-3}$, respectively~\cite{Workman:2022ynf}. In the $D^0 K^+$ invariant mass distributions of these process, there should be the signal of $\tcsbar^+$, which may be accessible for the LHCb Collaboration.


\section*{Acknowledgement}
This work is supported by the National Natural Science Foundation of China under Grant Nos. 11775050, 12175037, and 12192263. This work is also supported by the Natural Science Foundation of Henan under Grand Nos. 222300420554 and 232300421140,  the Project of Youth Backbone Teachers of Colleges and Universities of Henan Province (2020GGJS017), the Youth Talent Support Project of Henan (2021HYTP002), and the Open Project of Guangxi Key Laboratory of Nuclear Physics and Nuclear Technology, No.NLK2021-08.


\end{document}